\documentclass[prb,aps,twocolumn,groupedaddress,floats,showpacs,final, fleqn]{revtex4-1}
\usepackage{graphicx}
\usepackage{amsmath}
\usepackage{dcolumn}
\usepackage{bm}
\usepackage{color}
\usepackage{mathtools}
\usepackage{feynmp-auto}
\usepackage{hyperref}

\begin{document}
\definecolor{blue}{rgb}{0.3,0.3,0.9}

\author{Lode Pollet}
\affiliation{Department of Physics, Arnold Sommerfeld Center for Theoretical Physics, University of Munich, Theresienstrasse 37, 80333 Munich, Germany}
\affiliation{Wilczek Quantum Center, School of Physics and Astronomy and T. D. Lee Institute, Shanghai Jiao Tong University, Shanghai 200240, China}
\author{Nikolay V. Prokof'ev}
\affiliation{Department of Physics, University of Massachusetts, Amherst, MA 01003, USA}
\affiliation{National Research Center ``Kurchatov Institute,"
123182 Moscow, Russia}
\author{Boris V. Svistunov}
\affiliation{Department of Physics, University of Massachusetts,
Amherst, MA 01003, USA}
\affiliation{National Research Center ``Kurchatov Institute,"
123182 Moscow, Russia}
\affiliation{Wilczek Quantum Center, School of Physics and Astronomy and T. D. Lee Institute, Shanghai Jiao Tong University, Shanghai 200240, China}

\title{Stochastic lists: Sampling multi-variable functions with population methods}
\date{\today}
\begin{abstract}
We introduce the method of stochastic lists to deal with a multi-variable positive function, defined by a 
self-consistent equation, typical for certain problems in  physics and mathematics.
In this approach, the function's properties are represented statistically by lists containing a 
large collection of sets of coordinates (or ``walkers")
that are distributed according to the function's value. 
The coordinates are generated stochastically by the Metropolis algorithm and may replace older entries according to some protocol.
While stochastic lists offer a solution to the impossibility of efficiently computing and storing 
multi-variable functions without a systematic bias,
extrapolation in the inverse of the number of walkers is usually difficult, even though in practice very good results are found already for short lists.
This situation is reminiscent of diffusion Monte Carlo, and is hence generic for all population-based methods.
We illustrate the method by computing the lowest-order vertex corrections in Hedin's scheme for the Fr\"ohlich polaron and
the ground state energy and wavefunction of the Heisenberg model in two dimensions.
\end{abstract}

\maketitle

\section{Introduction}

We are interested in the solution $F(\mathbf{x})$ of equations of the type
\begin{equation}
F(\mathbf{x}) = F_0(\mathbf{x}) + K[F(\mathbf{x})],
\label{eq:selfcon}
\end{equation}
where the coordinate $\mathbf{x}$ is high-dimensional.  The (generically nonlinear) functional 
$K$ can involve a number of integrations,
multiplications and summations, but we do not consider differentiations.
Ultimately, the solution of (\ref{eq:selfcon}) is often used to compute integrals involving $F$ and some other, more simple,
functions. A straightforward approach to solve Eq.~(\ref{eq:selfcon}) is by fixed point iterations: Starting from a guess $F^{(0)}$,
one computes the right-hand side, plugs the newly obtained $F^{(1)}$ into the right-hand side, and continues this
iteration until, ideally, convergence is reached. However, as soon as the dimension is higher than three it becomes very
difficult to efficiently compute and store the function $F$.

Equations of the above type typically occur in the self-consistent formulation of quantum field theory, such as the Hedin
equations,\cite{Hedin1965} the Schwinger-Dyson equations,\cite{Dyson1949,Schwinger1951} parquet equations,\cite{parquet1,parquet2}
etc. They can also occur in the presence of spontaneous symmetry breaking, such as the Bardeen-Cooper-Schrieffer theory, when the
ordering field has to be determined self-consistently. Whereas the Green function, the self-energy, the polarization, and the effective
interaction are typically two-dimensional in case of rotational and translational symmetry [i.e., there is one spatial (or momentum)
coordinate and one time (or frequency) coordinate] and can be stored efficiently with standard grids, the irreducible three-point
vertex is already five-dimensional in 3D.
Studies attempting to solve the Hedin equations therefore often treat the vertex function as just a bare vertex, 
leading to the $GW$-approximation.
Alternatively, the full function is represented by an infinite number of contributions that only involve relatively simple integrals
[cf. the expansion of the Luttinger-Ward functional~\cite{Luttinger1960} in the Baym-Kadanoff effective action~\cite{Baym1961,Baym1962}
(or its generalization to bosons~\cite{DeDominicis1964a,DeDominicis1964b}) popular in the context of electronic structure calculations].

This problem---sometimes referred to as the curse of dimensions---is among the most prominent ones faced by diagrammatic Monte Carlo (DiagMC) methods.
This is unsatisfactory, because the premise of the DiagMC simulation is precisely to deal with high dimensions while maintaining
the central limit theorem. In this work, we introduce {\it stochastic lists}, a method in which the high-dimensional function $F(\mathbf{x})$
is represented by a stochastic list of coordinates $\mathbf{x}_1, \mathbf{x}_2, \ldots, \mathbf{x}_P$, with $P$ the length of the list.
The entries in the stochastic list are obtained by some Markov  process (to be discussed later; the key property is that the
values of $F(\mathbf{x})$ do not explicitly enter into the equation used for generation of the list entries), and can be refreshed
by some protocol, which is non-Markovian. The list provides a faithful statistical representation of $F(\mathbf{x})$: any quantity
of interest which can be written as some integral over $F$ can be computed. In this work we assume that $F$ is positive.

We benchmark the approach by considering two different systems. 
First, we study the Hedin equations for the Fr\"ohlich polaron
by computing the lowest order vertex corrections self-consistently. Second, we formulate the power method to find the ground state of a
Hamiltonian in the language of stochastic lists and study the ground-state energy and wavefunction of the two-dimensional Heisenberg model.
We find that, in practice, short lists give already remarkably accurate results. 
Typically, a power law extrapolation can be attempted over several decades in $1/P$. However, we found deviations for longer lists (in our examples when $P \gg 10^6$), which makes
any extrapolation difficult (unless the stochastic error dominates at this point).

As is clear from the power method example, stochastic lists share a number
of properties with the diffusion Monte Carlo method. An exponential scaling for diffusion Monte Carlo was reported previously in the
literature and was related to the correlation within the population of walkers as a consequence of the population control
mechanisms.\cite{Nemec2010}. It seems therefore that all population-based methods ultimately scale exponentially, implying that for
sufficiently large (and hard) systems the extrapolation cannot be done reliably to eliminate the systematic bias.
This outcome also questions the practicality of the method for arbitrary (bosonic) problems. Nevertheless, for many realistic
cases this scaling will not be seen (due to the dominance of the stochastic error), and by sufficient insight into the problem
(such as choosing a very good guiding wavefunction) the prefactor can be substantially reduced
such that the scaling is not an issue. This we can also demonstrate for stochastic lists.

This paper is structured as follows. In Section~\ref{sec:frohlich}, we study the non-crossing approximation and the
lowest non-trivial order vertex corrections for the Fr\"ohlich polaron problem. In Section~\ref{sec:heisenberg}, we proceed
with the ground state energy of the two-dimensional Heisenberg model, with a special emphasis on the numerical convergence
of the stochastic process. We conclude in Section~\ref{sec:conclusion}.

\section{The Hedin equations for the Fr{\"o}hlich polaron problem}\label{sec:frohlich}

As a first application, we consider the Hedin equations for the Fr\"ohlich polaron. This system has a positive expansion,
and is known to be convergent at any finite temperature. It is hence free from the most important restriction on the
DiagMC method, which is the series convergence. (Since the DiagMC algorithms work by iteration, they typically require
a finite region of convergence.) Sign-positive representation also implies that there is no need to take special care
of the diagram topologies, and one can proceed with standard Metropolis-Hastings sampling techniques. \cite{Metropolis, Hastings1970}
Therefore, the study of the Fr\"ohlich polaron provides an ideal opportunity to benchmark the idea of stochastic lists
in the context of vertex corrections.

\subsection{Model}\label{sec:model}

The Fr{\"o}hlich polaron model describes the interaction between an itinerant electron and longitudinal optical phonons
in insulators. Historically, it was the first problem to which the DiagMC method was applied~\cite{Prokofev1998, Mishchenko2000,Mishchenko2001}
and for which it was able to provide definite answers regarding the polaron spectrum and arbitrarily precise polaron energies
for any coupling strength. The Hamiltonian in the thermodynamic limit is given by ($\hbar=1$)
\begin{eqnarray}
H & = & H_{\rm el} + H_{\rm ph} + H_{\rm el-ph}  \\
H_{\rm el} & = & \int \frac{d^3 k}{(2 \pi )^3}  \frac{k^2}{2m} a_{\mathbf{k}}^{\dagger} a_{\mathbf{k}} \, ,
\nonumber \\
H_{\rm ph} & = &  \int \frac{d^3 q}{(2 \pi )^3} \omega_{\mathbf{q}} b_{\mathbf{q}}^{\dagger} b_{\mathbf{q}} = \omega_{\rm ph}  \int \frac{d^3 q}{(2 \pi )^3}  b_{\mathbf{q}}^{\dagger} b_{\mathbf{q}} \, ,\nonumber \\
H_{\rm el-ph} & = & \int \frac{d^3 k \, d^3 q }{(2 \pi )^6} \,  V(\mathbf{q})
(b_{\mathbf{q}}^{\dagger} -  b_{\mathbf{q}}) a_{\mathbf{k} - \mathbf{q}}^{\dagger} a_{\mathbf{k}} \, ,
\nonumber \\
V(\mathbf{q}) & = & \frac{ i \omega_{\rm ph}}{q \, (2m\omega_{\rm ph})^{1/4} } \left( \frac{4 \pi \alpha}{V} \right)^{1/2}
  =  \frac{i \tilde{\alpha}}{q}.
\nonumber
\end{eqnarray}
The operators $a_{\mathbf{k}}$ and $b_{\mathbf{q}}$ are annihilation operators for electrons of mass $m$ with momentum $\mathbf{k}$
and phonons with momentum $\mathbf{q}$, respectively. The phonon frequency $\omega_{\mathbf{q}} \equiv \omega_{\rm ph}$ can be taken
momentum-independent for optical, longitudinal phonons.
The dimensionless coupling constant is $\alpha$.  Typical values for $\alpha$ vary from $0.023$ for InSb over $0.29$ for CdTe
to $1.84$ for AgCl.\cite{Devreese2010}

Here we focus on the $T=0$ case.
In the imaginary-time representation, the bare propagator reads
$G_0(\mathbf{k}, \tau) =  - \theta(\tau) \exp \left[ - ( \frac{k^2}{2m} - \mu ) \tau \right] $.
The phonon propagator $D(\mathbf{q}, \tau) = (\tilde{\alpha} /q)^2 \exp( - \omega_{\rm ph} \tau)$
remains unrenormalized and is dispersionless. We absorbed the modulus squared of the electron-phonon interaction potential
into the phonon propagator for convenience (the two factors always enter the technique as a product).
The method of stochastic lists can not deal with momentum or frequency conservation because all coordinates of the
vertex function are generated by the list without restrictions.
We therefore need to work in the imaginary-time,  real-space representation, where
the propagators for phonons and electrons read $D(\mathbf{r}, \tau) = \frac{\tilde{\alpha}^2}{4 \pi r} \exp( - \omega_{\rm ph} \tau)$ and
$G_0(\mathbf{r}, \tau) = -\theta(\tau) \left( \frac{m}{2 \pi \tau} \right)^{3/2} \exp(- \frac{ m r^2}{2 \tau} + \mu \tau) $,
respectively.


\subsection{The Non-Crossing Approximation}
\label{sec:NCA}

\begin{figure}[tbp]
\centering
 \includegraphics[trim = 0 0mm 0mm 0, clip, width=0.6\columnwidth, angle=-90]{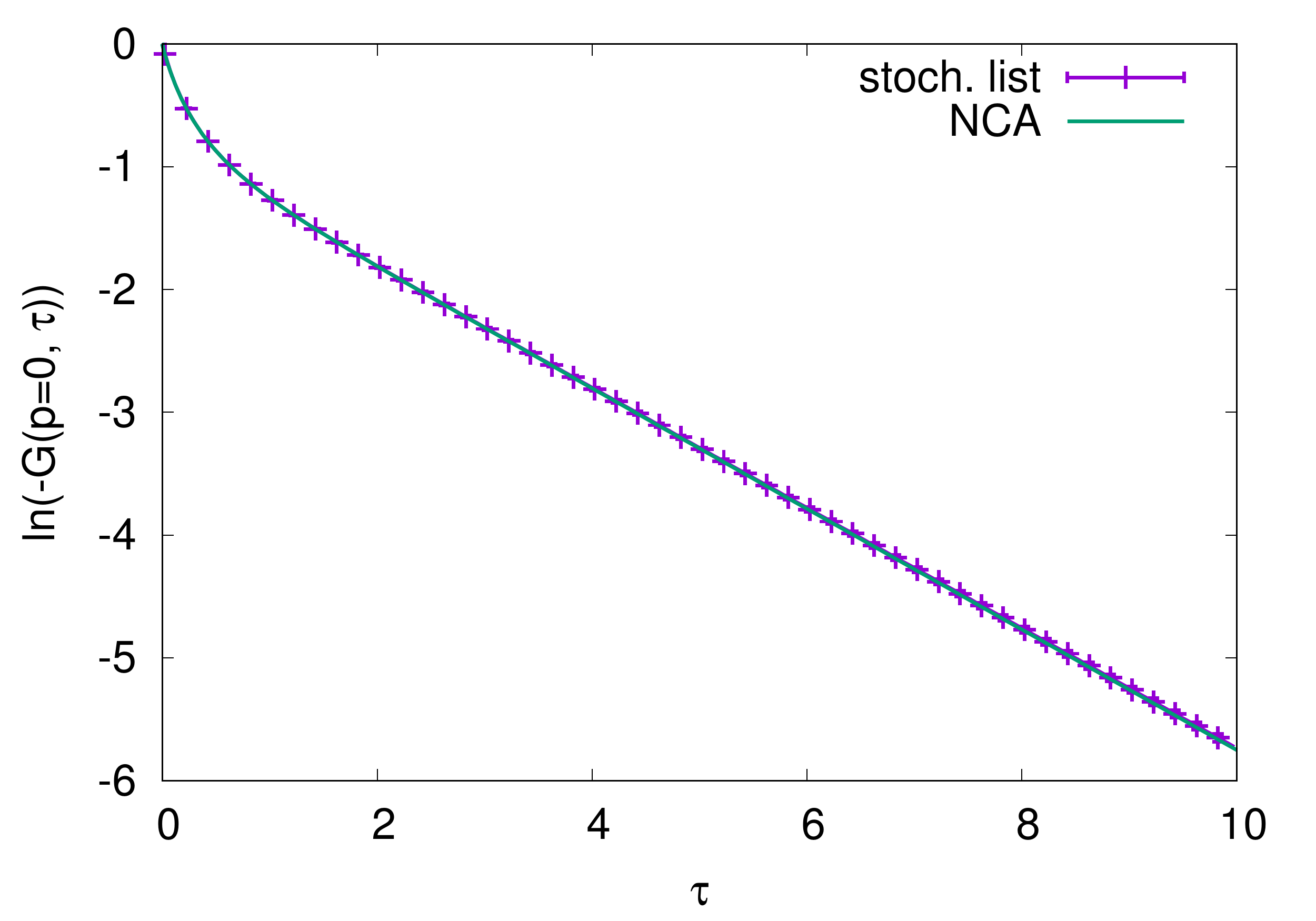}
\caption{(Color online.) Comparison of the Green function dependence on imaginary time at zero momentum, $-G(p=0, \tau)$,
obtained in the NCA approximation between the stochastic list method (stoch. list) and an explicit evaluation.
Data are shown for $\tilde{\alpha} = 5, \mu = -4$ and the list length $P=10^6$.}
\label{fig:nca_comp}
\end{figure}
As a first application to set the ideas, we solve the Fr\"ohlich polaron problem in the non-crossing approximation (NCA),
which corresponds to the first-order skeleton diagram. The coupled set of NCA equations reads
[we employ both $(\mathbf{k}, \omega_n)$ and $(\mathbf{r}, \tau)$ representations to simplify these equations]
\begin{eqnarray}
\Sigma^{(1)}(\mathbf{r}, \tau) &  = &  D(\mathbf{r}, \tau) \, G^{(1)}(\mathbf{r}, \tau) \, ,
\nonumber \\
G^{(1)}(\mathbf{k}, \omega_n) & = & \frac{1}{G_0^{-1}(\mathbf{k}, \omega_n)  - \Sigma^{(1)} (\mathbf{k}, \omega_n)  }.
\end{eqnarray}
Reference NCA data can be found in the lecture notes Ref.~\onlinecite{Frohlich_lecturenotes}.

In order to solve these equations with stochastic lists, we cast the entire setup as a single
self-consistent non-linear integral equation of the form (\ref{eq:selfcon}):
\begin{eqnarray}
G^{(1)}(X) =  G_0(X)  +  \int {\cal I}\, d^4 X_1 \, d^4 X_2 ,\qquad  \qquad \label{eq:nca_list}  \\
 {\cal I}  = G_0(X_1) \, G^{(1)}(X_2 \! -\!  X_1) \, D(X_2\! -\! X_1)\,  G^{(1)}( X\!  - \! X_2 ), \nonumber
\end{eqnarray}
where we introduced the 4-dimensional space-time position vectors $X=(\mathbf{r}, \tau)$, $X_1 = (\mathbf{r}_1, \tau_1)$,
and $X_2 = (\mathbf{r}_2, \tau_2)$. We then pretend that $G^{(1)}(X)$ cannot be evaluated and stored as a function, but its properties can be represented by
a collection---the list---of $X$-coordinates generated stochastically by sampling the r.h.s. of (\ref{eq:nca_list}) using standard
DiagMC techniques. There is no gain in using rotational symmetry for the coordinates in the list.

The minimal set of updates consists of switching between the two sectors corresponding to the first and second terms
in the r.h.s. of (\ref{eq:nca_list}). The known integral over the first term,
${\cal N}_0= \int \vert G_0(X) \vert d^4X = 1/\vert \mu \vert $,
is used for normalization: the Monte Carlo statistics for any property is properly normalized
once multiplied by the factor ${\cal N}_0/Z_1$, where $Z_1$ is the number of samples that belong to the
first term. For example, the normalization integral
${\cal N}_G = \int \vert G^{(1)} (X) \vert d^4X$ is obtained as
${\cal N}_G = {\cal N}_0 (Z/Z_1)$, where $Z$ is the total number of samples (no matter whether they belong
to the first or the second term).

In order to go from the first to the second term, we draw a random variable $\tau_1$ according to an exponential
distribution $\propto \vert \mu \vert e^{- \vert \mu \vert \tau_1} d\tau_1$ and then generate
 three random numbers $(x_1, y_1, z_1)$ according to a gaussian distribution with zero mean and variance $\tau_1/m$;
the corresponding probability density is based on the $G_0$ function.
Next, we choose coordinates $X_3$ and $X_4$ uniformly from the existing list; they define $X_2=X_1+X_3$ and $X=X_2+X_4$,
by translation invariance. The probability $1/P^2$ for selecting these two variables from the list
is a faithful representation of the probability $G^{(1)}(X_2 \! -\!  X_1) \, G^{(1)}( X\!  - \! X_2 ) \, d^4X_2 \, d^4 X / {\cal N}_G^2$.
In the reverse update taking the simulation from the second to the first term,
the coordinates of the free propagator are again determined by the probability density based on $G_0$:
an exponential random number for the time and three gaussian random numbers for the space.

The key property behind the list technique is that all the values of the unknown function $G^{(1)}$ cancel in the acceptance ratio: The unknown full Green functions appear in both the proposed configuration weight and the probability density
used to generate new variables. For the same reason all exponential and gaussian factors cancel $G_0(X_1)$. This results in an acceptance ratio $R = D(X_2\! -\! X_1) \, {\cal N}_G^2 $ (and $1/R$ for the reverse update).
The update is then accepted with probability ${\rm min}(1,R)$, according to the Metropolis-Hastings algorithm.
One thus arrives at a protocol of dealing with a function without knowing/revealing its explicit form.

In this implementation, we work with an existing list from which we draw random coordinates, while simultaneously preparing
a new list which we consecutively fill after each Monte Carlo update by recording the current value of $X$
(in both sectors). When the new list is full, it replaces the existing one and the new list is reset to zero.
To initialize the procedure, we start with a short list based on random coordinates that we draw from the $G_0$ distribution,
which we let grow by a small factor of the order of $1.001 \div 1.01$ till the maximum length is reached.

The results for a moderate coupling $\tilde{\alpha}=5$ are shown in Fig.~\ref{fig:nca_comp}.
Within the level of resolution of the plot, there is no difference between the exact NCA result and the one obtained
by stochastic lists of the length $P = 10^6$. 
We postpone the discussion of convergence properties till Sec.~\ref{sec:heisenberg}.
Here we simply note that for $P = 10^6$ any systematic bias was subdominant to the statistical noise. Also,
there is no need for fast Fourier transforms 
(cf. Ref.~\onlinecite{Frohlich_lecturenotes}) when employing lists
and thus no need to take care of the asymptotic behavior of the Green function for large frequencies.

\subsection{First-order vertex corrections}


\begin{figure}[tbp]
\centering
 \includegraphics[trim = 0 0mm 0mm 0, clip, width=\columnwidth, angle=0]{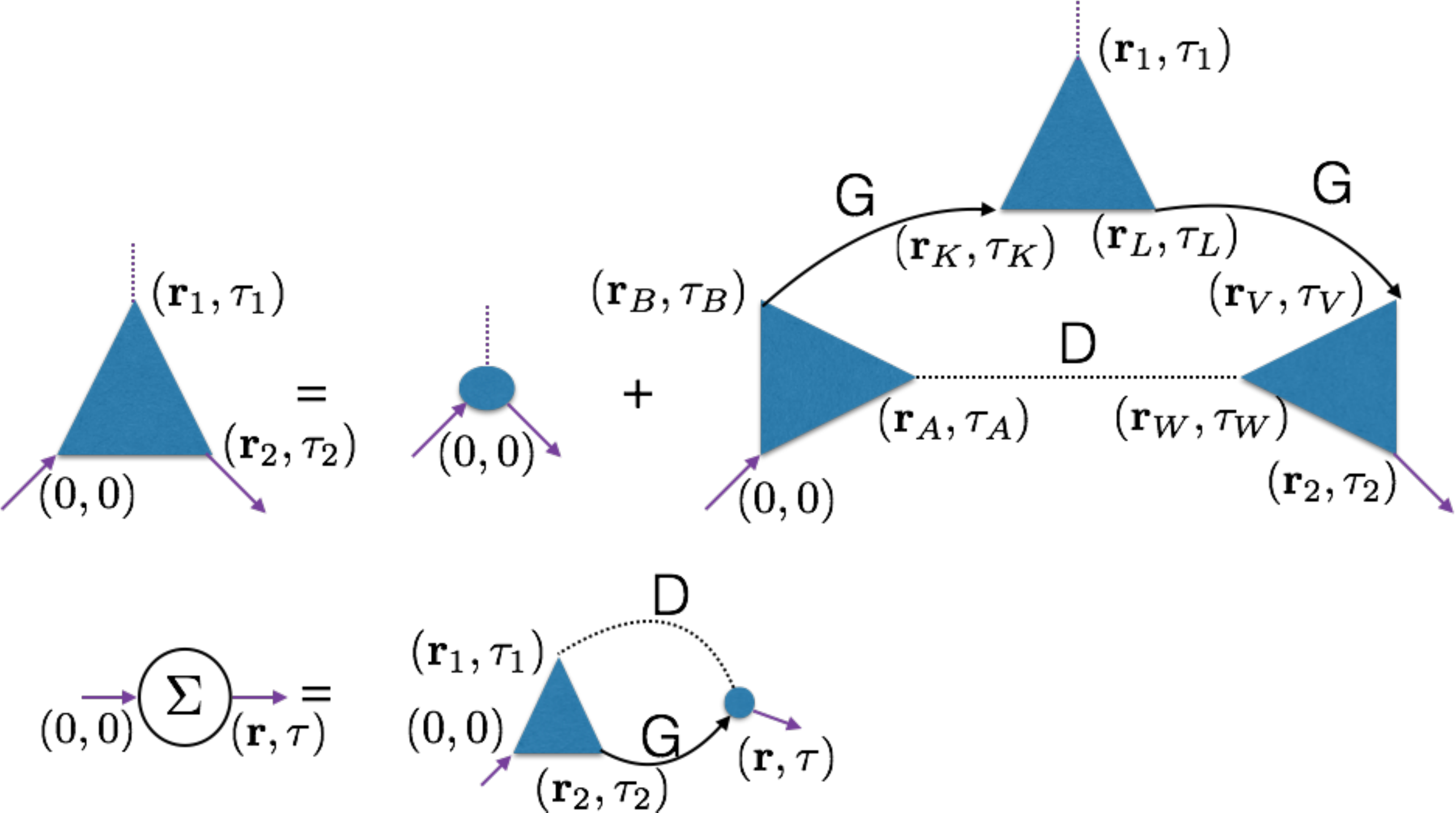}
\caption{(Color online.) {\it Upper pane}: Lowest-order contributions to the three-point vertex in the (imaginary-time, real-space) representation.  {\it Lower pane}:  Self-energy in terms of the three-point vertex in the (imaginary-time, real-space) representation. }
\label{fig:3p_vertex}
\end{figure}

\begin{figure}[tbp]
\centering
 \includegraphics[trim = 0 0mm 0mm 0, clip, width=1.0\columnwidth]{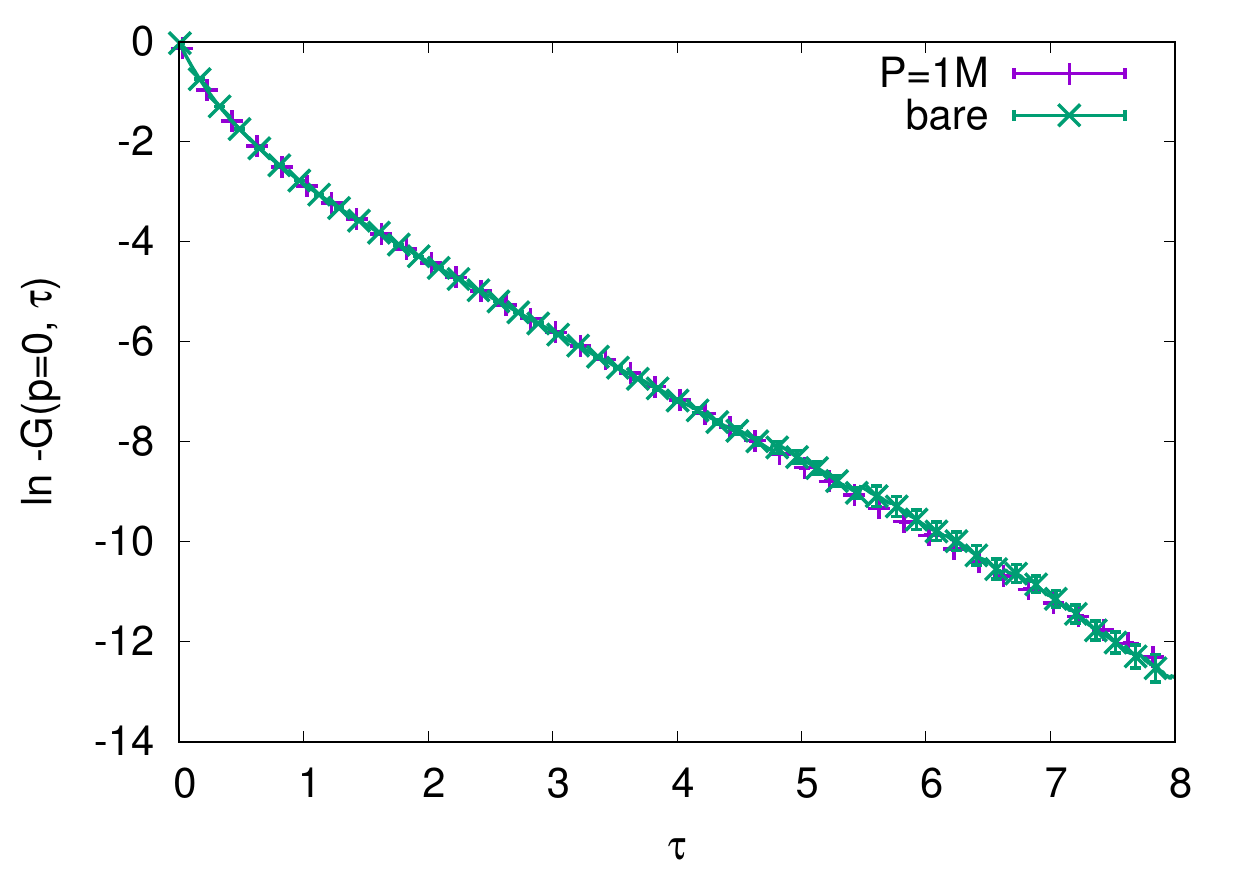}
\caption{(Color online.) Comparison of the Green function dependence on imaginary time at zero momentum, $-G(p=0, \tau)$,
obtained by the stochastic list method applied to the scheme illustrated in
Fig.~\ref{fig:3p_vertex} with a list of length $P=10^6$ (``P=1M")
versus the result obtained by a conventional DiagMC simulation restricted to sample the same set of diagrams from the
bare series (see text).}
\label{fig:comparison_vertex}
\end{figure}

We now consider the Hedin scheme~\cite{Hedin1965} where the vertex function takes into account both the zeroth-order term and the first-order corrections.
The system to solve consists of three equations, the two of which are shown graphically in Fig.~\ref{fig:3p_vertex},
and the third one is the Dyson equation for the Green function.
Our implementation involves two stochastic lists: one for the Green function $G(X)$, as before,
and the other one for the three-point vertex $\Gamma(X_1,X_2)$, the latter containing 6 spatial and 2 temporal coordinates.
(As before, there is no gain in exploiting symmetries to reduce the number of spatial coordinates.)
We set up one stochastic Markov-chain process to sample both quantities.
The algorithm itself is a straightforward extension of the one discussed above, and we will not elaborate here on minute technical details.

Reference data were obtained from the algorithms discussed in Ref.~\onlinecite{Frohlich_lecturenotes},
where we used the bare-series  code with the updates ``insert-remove" and ``dress-undress" switched on and the
``swap" update switched off. Starting from an arbitrary Green function diagram, the ``insert" update attempts
to insert a new $D$-propagator without dressing any of the existing vertices; i.e., none
of the vertices covered by the new $D$-propagator remains unlinked on the updated time interval.
``Remove" is the complementary update.
The ``dress" update attempts to insert a new $D$-propagator that covers precisely one vertex;
the ``undress" update is its complementary partner.
This set of updates is not ergodic for the full problem (e.g., no diagrams in which a $D$-propagator
covers two or more non-linked vertices can be reached),
but it accounts for all diagrams covered by Fig.~\ref{fig:3p_vertex}.

The results of  benchmark comparison are shown in Fig.~\ref{fig:comparison_vertex}.
We see that perfect agreement is reached for the Green function at zero momentum as a function of imaginary time
for a list of length $P=10^6$ (the same length is used for both $G$ and $\Gamma$).
The method of stochastic lists seems thus promising to study vertex corrections in the context of (bosonic)
dynamical mean-field theory and its cluster extensions. However, as we will see in the next section,
establishing the convergence of the answer can only be done on a case by case basis, at best.

\section{Ground state of the antiferromagnetic Heisenberg model}\label{sec:heisenberg}


\begin{figure}[tbp]
\centering
 \includegraphics[trim = 0 0mm 0mm 0, clip, angle=-90, width=1.0\columnwidth]{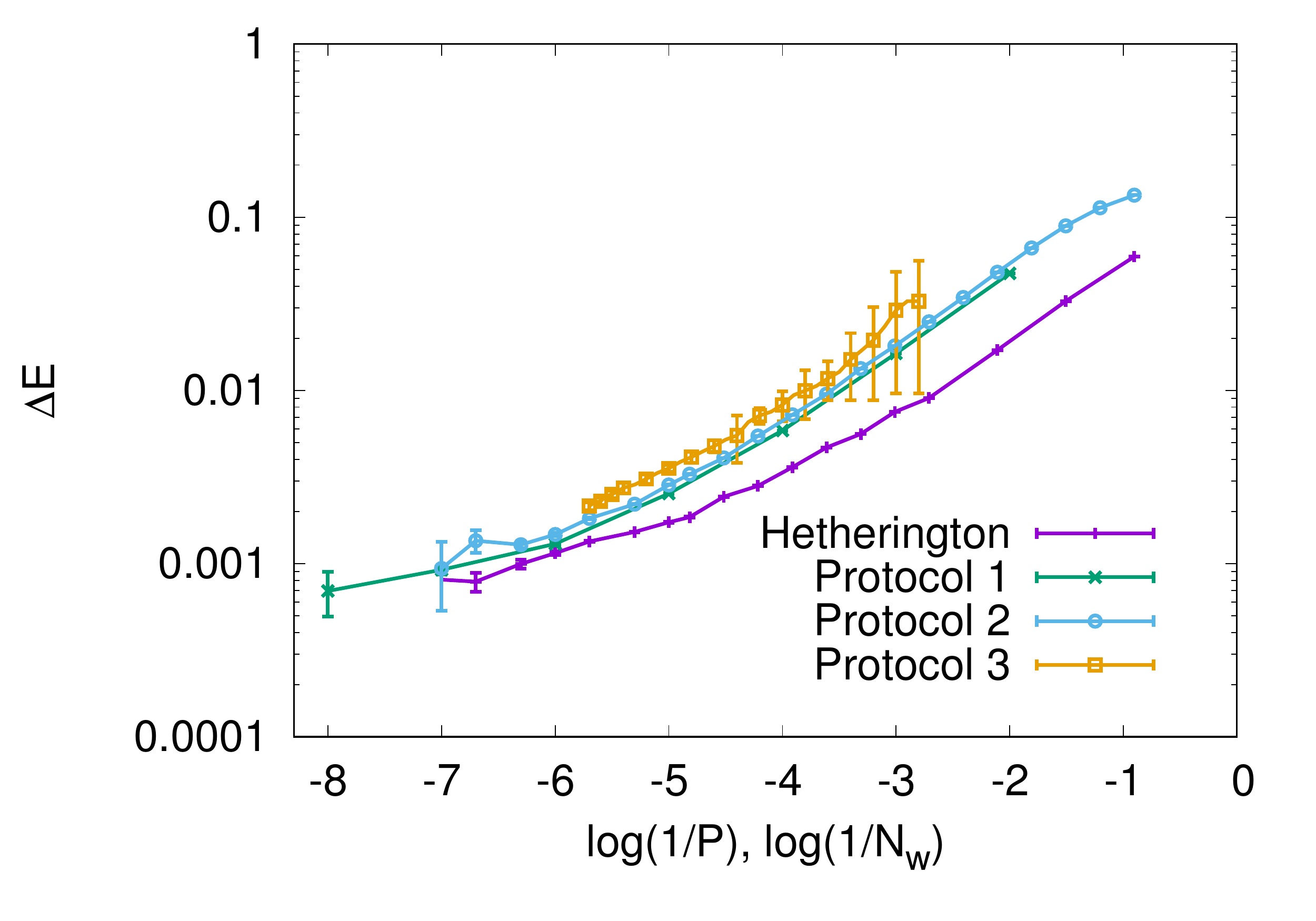}
\caption{(Color online.) Difference between the exact energy per site (taken from Ref.~\onlinecite{Sandvik1997}) and the one obtained
with either a stochastic list of length $P$ (see text for an explanation of the different protocols), or in a diffusion Monte Carlo simulation with $N_w$ walkers
(``Hetherington," with $k=128$; see text). The system is a 2D Heisenberg model with linear size $L=10$ with $J=1$.
}
\label{fig:scaling}
\end{figure}


\begin{figure}[tbp]
\centering
 \includegraphics[trim = 0 0mm 0mm 0, clip, angle=-90, width=\columnwidth]{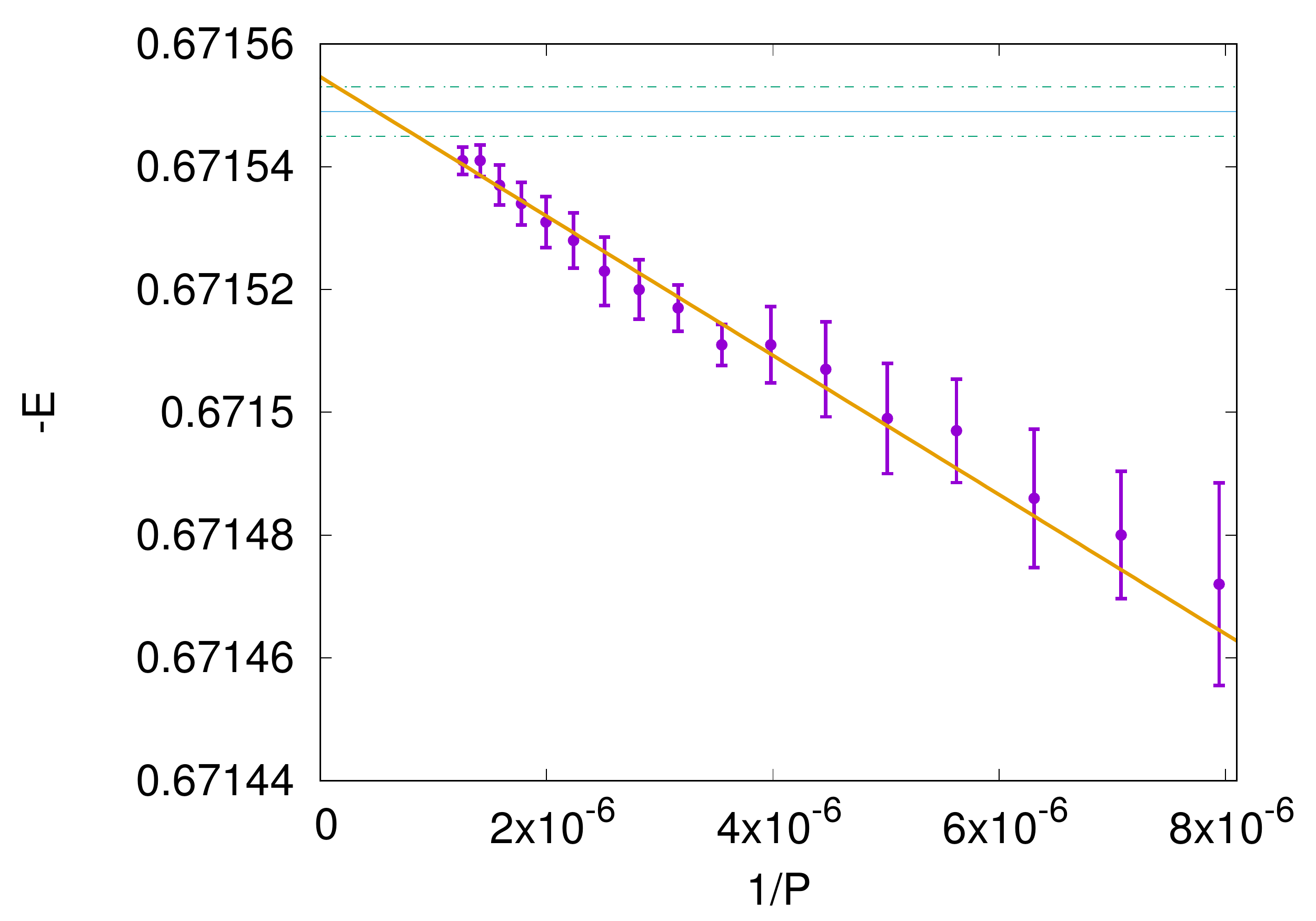}
\caption{(Color online.) Minus the energy per site  as a function of $1/P$ using protocol 3 (with $\kappa = 1$) with a Gutzwiller guiding wavefunction with $b=0.8$, which is (close to) optimal. The error bars have been obtained from 10 independent runs. The data are compared with the  value $E/J = - 0.671549(4)$ per spin, plotted as a thin blue line and its error bars as a thin dot-dashed green line, obtained by A. Sandvik using the stochastic series expansion method~\cite{Sandvik1997}. 
The data have been extrapolated linearly (shown as a full line in the plot) according to $E(x) = E_0 - b x$ with $x = 1/P$, resulting in $a = 0.671555(4)$ and $b = 11.3(4)$.
The system is a 2D Heisenberg model with linear size $L=10$ with $J=1$.
}
\label{fig:guiding}
\end{figure}

In this section, we consider the spin-$1/2$  Heisenberg antiferromagnet (HAF) on a square lattice
%
%
\begin{equation}
H  = J \sum_{\left< i,j \right>} \mathbf{S}_i \cdot \mathbf{S}_j
 =  {J\over 2} \sum_{\left< i,j \right>}  ( S_i^{+} S_j^{-} + S_i^{-} S_j^{+} + 2S_i^z S_j^z),
\end{equation}
with spin exchange amplitude $J > 0$. The sum is over nearest neighbor sites, and the lattice is of size $L\times L$.
By performing the unitary transformation $S_i^x \to -S_i^x, \quad S_i^y \to -S_i^y, \quad S_i^z \to S_i^z$ on one of
the sublattices, the sign of the amplitude in front of the raising and lowering term is reversed. The matrix elements
of $C - H$, for an appropriately chosen constant $C = J L^2 /2$, are then all positive in the usual $S^z$ basis.
The eigenvalue problem
\begin{equation}
(C - H) \psi = (C - E_0) \psi,
\end{equation}
can be considered a power method when the state $\psi$ is iteratively represented by the stochastic list.
It yields the absolute value of the largest eigenvalue in magnitude, whose eigenfunction
can always be chosen positive for a positive matrix. With the above transformations, this corresponds
to projecting onto the antiferromagnetic ground state (and not the ferromagnetic anti-groundstate).
This problem is challenging because of the gapless spectrum of elementary excitations in the thermodynamic
limit and the large dimension of the Hilbert space, growing  exponentially with system size.

The coordinates in the list consist now of $L^2$ bits representing the spins on the lattice.
We add to the sampled configuration space a dummy spin-independent term for normalization purposes; the
equation to solve thus contains a normalization constant $C_N$ as well as the matrix-vector multiplication term.
The key updates are switching between these two terms.
To go from the former to the latter, we pick randomly an entry from the stochastic list of length $P$,
which provides us a coordinate (Fock state) $j$. The probability of this selection is $1/P$ or
$\vert \psi_j \vert / {\cal N}_{\psi}$, where  ${\cal N}_{\psi} =\sum_j \vert \psi_j \vert$ is the
normalization sum (estimated using the same procedure as described above for ${\cal N}_G$).
Next, we have to determine all non-zero Hamilton matrix elements  $H_{ij}$ when acting on the state
corresponding to coordinate $j$. We assume that the Hamiltonian is too large to be stored explicitly,
so that this step must be repeated in every Monte Carlo update.  For sparse matrices, there are very
few possible final coordinates $i$. We choose the final coordinate $i$ according to the heatbath algorithm,
i.e. with probability $ p_i =  H_{ij} /\bar{H}_j$, where $\bar{H}_j = \sum_i  H_{ij} $. The resulting Metropolis-Hastings acceptance ratio is just $R = {\cal N}_{\psi} \bar{H}_j / C_N $
(and $1/R$ for the reverse update). The update is accepted with probability $\rm{min}(1,R)$.
We also occasionally employed the Metropolis-like algorithm, in which one of the non-zero matrix elements $H_{ij}$
is chosen with uniform probability instead of the heatbath algorithm. It did not lead to significant differences in the autocorrelation times.
A good choice for the dummy term constant is $C_N = 2 J L^2$ (to compensate the typical value of the
$\bar{H}_j$ sum). Measurements of the  parameter ${\cal N}_{\psi}$ and recordings of the new list
entries are performed only in the matrix-vector multiplication sector.

Apart from the previously discussed protocol (``protocol 1") where the current list is replaced by the
new one once the latter is completed, we also applied ``protocol 2." Here there is only one list updated
at each Monte Carlo step by drawing a random integer in the interval $m \in [0,P[$
and replacing the existing entry $m$ with the new coordinate.
In ``protocol 3" the list is continuously growing as a function of the Monte Carlo steps according to $P = \sqrt{\tau_{\rm MC} / \kappa}$.
The list grows then by one entry whenever the integer part of $\sqrt{\tau_{\rm MC} / \kappa}$ increases by one; otherwise an existing entry is overwritten when measuring the list.
As will be clear from the results, the differences between these protocols are not of leading importance.

The above algorithm works remarkably well for small matrices, even if they are poorly conditioned.
For a system of size $L=4$, the systematic error can easily be made smaller than the statistical noise.
When $P$ exceeds 1000, we find that the systematic error decreases as $1/P$. However, at larger
system size (speaking of $L=10$ and larger), this extrapolation law is no longer valid: Longer lists are needed, which need to be iterated much longer to converge.
We also observe that the converged results cannot be extrapolated as a power law in $1/P$ (even though the results are remarkably accurate already for short lists).  
This gets worse with increasing system size, and the marginal gain of using longer lists diminishes further. We also see in
Fig.~\ref{fig:scaling} that the scaling does not depend on which protocol we use, suggesting that the
reason for the inefficiency of the simulation must be found in the build-up of autocorrelations scaling
unfavorably with the system size: due to the overwriting of the entries in the stochastic list, a few
Fock states tend to dominate and the superposition of those is not the exact ground state.
Protocol 3 appears to yield results that can be extrapolated by a single power law over several decades in $1/P$ and may hence look superior. Some care with this observation is 
however needed because we could not reach lists that are as long as in protocols 1 and 2;  {\it i.e.,} it might be that the law $P = \sqrt{\tau_{\rm MC} / \kappa}$
is still too fast when $P \gg 10^5$.

The method of stochastic lists applied to the power method is highly reminiscent of diffusion Monte Carlo,
which has also been applied to the Heisenberg model with impressive
results.\cite{Hetherington1984,Trivedi1990,Runge1992a,Runge1992b,Umrigar1993,Sorella1998}
We compare here with an implementation motivated by the original algorithm by Hetherington~\cite{Hetherington1984} and
the reconfiguration ideas of Ref.~\onlinecite{Sorella1998}. In this scheme, $N_w$ walkers propagate through the Fock space but,
instead of satisfying a detailed balance condition, they acquire multiplicative weight factors (these are the previously
introduced $\bar{H}_j$ sums), which fluctuate at an extensive scale. After a number of generations $k$, a population-control
mechanism is applied to keep the number of walkers fixed. Walkers with high weight are more likely to reproduce and walkers
with a low weight are more likely to be eliminated. The resulting bias is compensated by global factors
$\left< \bar{H}(1) \right>, \ldots \, , \left< \bar{H}(k) \right>$ (i.e., the $\bar{H}$ values averaged over all
walkers in every generation), see Refs.~\onlinecite{Hetherington1984,Sorella1998} for details.
In Fig.~\ref{fig:scaling} we see that the above algorithm with $k=128$ yields results that are
more accurate than the list when there are few walkers, but that the scaling is the same as for the list.
We checked that the same holds for $k=12$. It has been known since the early days of diffusion Monte Carlo
that the population size might easily lead to the dominant source of error; more recently, Nemec
claimed an exponential scaling,\cite{Nemec2010} and population size bias was also found by Boninsegni and Moroni.\cite{Boninsegni2012}
The explanation given by Nemec apparently also applies to stochastic lists.

It is well known that diffusion Monte Carlo can significantly be enhanced by using a good guiding wavefunction.
For the HAF model, and certainly for small system sizes, excellent variational Jastrow wavefunctions are known.\cite{Franjic1997}
We employed here a simpler but faster to evaluate Gutzwiller ansatz, reminiscent of perturbation theory,
\begin{equation}
\psi_G \sim \prod_{\langle i,j \rangle}  \exp( -b S_i^z S_j^z),
\label{eq:guidingwvf}
\end{equation}
where $b$ is a variational parameter (Note however that we have no proof that our final answer for the ground state energy is variational). For $b > 0$ antiferromagnetic correlations are enhanced. 
The only change to the code is that the Hamiltonian matrix element is replaced by $H_{ij} = \psi_G(i) H_{ij} /  \psi_G(j)$ where $\psi_G(i)$ denotes Eq.~\ref{eq:guidingwvf} evaluated for spin configuration $i$.
We performed the simulation for various values of $b$ using protocol 3. We show in Fig.~\ref{fig:guiding} the convergence for $b = 0.8$, which is very close to optimal.  We find that the energies differ by an amount $ 8 \times \sim 10^{-6}$ for the longest lists we have studied, $P \sim 8 \times 10^5$. The figure makes however clear that the energy still drifts as a function of $1/P$. If we extrapolate, the results agree within error bars (of the order of $4 \times 10^{-6}$) with Sandvik's stochastic series expansion results~\cite{Sandvik1997} and with the diffusion Monte Carlo results of Ref.~\onlinecite{Sorella1998}.These results are hence up to two orders of magnitude more precise than the results
without using the guiding wavefunction. However, a poor guiding wavefunction can lead to severe slowing down and, recalling our main goal of studying vertex corrections, one has to recognize that good (and easy to evaluate) guiding schemes
are not available in general. 

\section{Conclusion}
\label{sec:conclusion}

We have introduced the method of stochastic lists, which allows one to accurately emulate properties
of a  multi-variable function $F(\mathbf{x})$. The repeatedly refreshed list consists of a large set of coordinates $\mathbf{x}$
distributed according to the $F(\mathbf{x})$ values. The list of length $P$
represents only a tiny fraction of the full coordinate space of $F$, but after it is refreshed multiple times,
a faithful representation of the entire $F(\mathbf{x})$ function is obtained.
The method was benchmarked by computing vertex corrections self-consistently for the Fr\"ohlich polaron model
and by applying the power method for obtaining the ground-state energy and wave function of the antiferromagnetic
Heisenberg model. 
The method gives reasonably accurate results for most problems in practice,
and can apparently be extrapolated as a power law over several decades in the inverse list length. 
However, for very long lists we could observe deviations rendering a controlled extrapolation for an arbitrary problem difficult.
This behavior seems inherent to all population-based methods.
Nevertheless, stochastic lists are extremely promising
for many problems where clever guiding functions for the stochastic sampling are known. Here the systematic
error can be reduced to such a degree that it becomes irrelevant in practice.

{\it Acknowledgement.}  We wish to thank M. Boninsegni for valuable discussions.
This work was supported by H2020/ERC Consolidator Grant No. 771891 (QSIMCORR),
the Munich Quantum Center and the DFG through Nano-Initiative Munich, the Simons Collaboration on the Many Electron Problem,
and the National Science Foundation under the grant DMR-1720465.
The open data for this project , including an open source implementation for Fig.~\ref{fig:guiding}, can be found at \url{https://gitlab.lrz.de/QSIMCORR/StochasticList}.  

\bibliographystyle{apsrev4-1}
\bibliography{refs}

\end{document}